%Paper: hep-th/9307110
%From: Yaroslav P. Pugay <pugay@itp.sherna.msk.su>
%Date: Sat, 17 Jul 1993 17:29:07 +0400

\input vanilla.sty
\font\tenbf=cmbx10

\font\ninebf=cmbx9
\font\ninerm=cmr9
\font\nineit=cmti9

\font\eightrm=cmr8
\font\eightit=cmti8

\TagsOnRight
\hsize=5.0truein
\vsize=7.8truein
\parindent=15pt
\baselineskip=10pt
\def\qed{\hbox{${\vcenter{\vbox{
    \hrule height 0.4pt\hbox{\vrule width 0.4pt height 6pt
    \kern5pt\vrule width 0.4pt}\hrule height 0.4pt}}}$}}
%%%%%%%%%%%%%%%%%%%%%%%%%%%%%%%%%%%%%%%%%%%%%%%%%%%%%%%%%%%%%%%%%%%%%%%%%%%%%
% def of CENTERCAPT with width 4.0 in for the centered multiline captures

%%%%%%%%%%%%%%%%%%%%%%%%%%%%%%%%%%%%%%%%%%%%%%%%%%%%%%%%%%%%%%%%%%%%%%%%%%%%%%
\line{\eightrm  LANDAU-TMP-1-93\hfil}
\vglue 5pc
\baselineskip=13pt
\line{\eightrm  hep-th/9307110\hfil}
\vglue 5pc
\baselineskip=13pt
%%%%%%%%%%%%%%%%%%%%%%%%%%%%%%%%%%%%%%%%%%%%%%%%%%%%%%%%%%%%%%%%%%%%%%%%%%
\centerline{\tenbf  NOTES ON $WGL_n$-ALGEBRAS AND}
\centerline{\tenbf  QUANTUM MIURA TRANSFORMATION.}
%%%%%%%%%%%%%%%%%%%%%%%%%%%%%%%%%%%%%%%%%%%%%%%%%%%%%%%%%%%%%%%%%%%%%%%%%%%
\vglue 24pt
%%%%%%%%%%%%%%%%%%%%%%%%%%%%%%%%%%%%%%%%%%%%%%%%%%%%%%%%%%%%%%%%%%%%%%%%%%%
\centerline{\eightrm Ya.P. Pugay
%%%%%%%%%%%%%%%%%%%%%%%%%%%%%%%%%%%%%%%%%%%%%%%%%%%%%%%%%%%%%%%%%%%%%%%%%%%
\footnote"$^{+}$"
{\eightrm\baselineskip=10pt supported by Landau Schoolarship awarded
by Forkschurgzentrum Julich GmbH and in part by the Soros Foundation
Grant of the American Physical society.
}
}
\baselineskip=12pt
\centerline{\eightit Landau Institute for Theoretical Physics}
\baselineskip=10pt
\centerline{\eightit 142432 Chernogolovka, Russia}
\vglue 16pt
\centerline{\eightrm ABSTRACT}
{\rightskip=1.5pc
\leftskip=1.5pc
\eightrm\baselineskip=10pt\parindent=1pc
We start from the quantum Miura transformation [7]
for the $W$-algebra associated with $GL(n)$ group
and find an evident formula for quantum
L-operator as well as for the action of $W_l$ currents (l=1,..,n)
on elements of the completely degenerated
n-dimensional representation. Quantum formulae are
obtained through the deformation of the pseudodifferential
symbols. This deformation is independent of $n$ and preserves
Adler's trace. Our main instrument of the proof is
the notation of pseudodifferential symbol with right action
which has no counterpart in classical theory.
\vglue 5pt
\baselineskip=13pt
\line{\tenbf 1. Introduction \hfil}
\vglue 5pt
W-algebras origionally introduced by Zamolodchikov [1] as a
generalization of the Virasoro algebra in context
of two-dimensional Conformal Field Theory [4] turned out to
be extremely interesting object. In the past few years
considerable progress has been made in an understanding
of the deep structures underlying these algebras
(see for example refs.[2,5-15]) as well as its classical
limits [17-20,24,25].

In this work we want to consider an aspect of the
W-algebras (associated with general linear group)
which seem to have been overlooked in the
literature.
In spite of the number of results reffering to
the $W_n$-algebras with an arbitrary $n$ several
principal questions still remain to be answered.
It is well-known that classical $W$-algebras
associated with general linear group
are isomorphic to the Poisson algebras of
functionals on the manifold of the linear
differential operators (with second Gel'fand-Dikii
bracket, as Poisson bracket) [3]. Other $W$-algebras
($WA_n$, $WB_n$, $WC_n$ ,$WD_n$) can be considered
as certain reductions of $WGL$-algebra [16]. This representation
in terms of (pseudo)differential operators is very
usefull.
However, at first glance it would seem that classical
theory loses many of its attractive features under quantization .
The reason is the lack of any suitable for work
formulae for L-operator, Lie brackets of the
generators etc.

Developing the work by Lukyanov
and Fateev [6] we show that there is exist
simple deformation of (pseudo)differential
operators which reproduces formulae for
quantum Miura transformation, L-operator
and the action of W-algebra on some invariant
operators (from the kernel of
L-operator). We hope that in the case of $W$-algebras
associated with other then $GL$-group [8] the similar
formulae still exist.

The paper is organized as follows. In sect.2
we review some necessary facts from classical
theory and derive an explicit formula
for $W_n$ algebra action on the Bloch's
solutions of the equation
$$
L\Psi(x)=\bigl(\sum_{i=1}^n W_i(x){\partial^{n-i}}\bigr)\Psi(x)=0.
\tag 1.1$$
Our calculation of well-known formula (2.7) in terms of Darboux
variables admit simple quantum generalization.
In sect.3 we disscuss the main difficulties appeared in the
quantum case and give the direct calculation
of the quantum $W_l$ ($l=2,3,4$) currents action on the highest
weight operator $F_1=:\exp(\phi_1):$ of the n-dimensional
completely degenerated representation
of $WGL_n$. We represent explicite formulae
for quantum L-operator for $n=2,3,4$ cases.
In sect.4 we introduce the formalism of
right derivations and bi-pseudodifferential operators.
Quantum Miura transformation is considered then as
the equality of two bi-differential operators.
We propose some deformation of the
(pseudo)differential operators $Y$ and $L$ preserving
Adler's trace of the product $YL$. In sect.5 we give general proof
for an arbitrary $n$ that such deformations of
the differential operators correspond to the quantum
$W_n$-algebras.
We work essentially with highest weight operator but all
formulae are valid for any of the $n$ invariant operators
obtained by the action of screening operators on $F_1$ such as
laters represent nilpotent part of
quantum group $U_q(gl(n))$ which commutes with an
current algebra [11,12].

We finally remark that some of the results of this paper
were published (without proof) in [23].
%\headline={\ifodd\pageno\rightheadline \else \leftheadline\fi}
%\def\rightheadline{\eightit \hfil International Journal
%of Modern Physics A \hfil\eightrm\folio}
%\def\leftheadline{\eightrm\folio\hfil \eightit Notes on $WGL_n$ algebras
%\hfil}
\voffset=2\baselineskip
\vglue 12pt
\line{\tenbf 2.CLASSICAL CASE. \hfil}
\vglue 5pt
Let us consider $WGL(n)$-invariant classical 2D field theory
on the circle.
Introduce scalar fields (Darboux variables) $\phi^{b}$ with expansion
$$
\phi^{b}(x)=q^{b}+p^{b}{\ln(x) \over {2{\pi}i}}+\sum_{m\neq 0}
{{{a_{-m}}^{b}x^m}\over {im}}   \quad ,
\tag 2.1$$
$x=\exp(i\sigma)$;$\sigma \in [\sigma_{0}$,
$\sigma_{0}+2\pi]$;$b=1$,$\ldots,n$
and Poisson brackets:
$$
\{\phi^{a}(x),\phi^{b}(y)\}={\pi}i sign(x-y)\delta^{ab}
$$
$$
\{a_{m}^{b},a_{n}^{c}\}=n\delta_{m+n,0}\delta^{bc},\{q^{a},p^{b}\}=
2{\pi}i\delta^{ab},\{q^{a},a_{m}^{b}\}=\{p^{a},a_{m}^{b}\}=0 \quad .
\tag 2.2$$
Then classical $W$-currents
are expressed via Miura transformation which for the groups
$G=GL(n)$,$(SL(n))$ has the form [3,16]:
$$
L(x)=\partial^{n}+\sum^{n}_{l=1} W_{l}(x) \partial^{n-l}=
\prod^{n}_{i=1}(\partial-\vec{h}_i\vec{\phi}'(x))
\tag 2.3$$
where $\vec{\alpha}_i$ and
$\vec{h}_{i}=\vec{\omega}_{1}- \vec{\alpha}_{1}-\ldots-
\vec{\alpha}_{i-1}$ are correspondently positive
roots of the Lie algebra $G$ and fundamental weights of the $n$-dimensional
vector representation of $GL(n)$,$SL(n)$. If we consider $n$-order
linear differential equation
$$
L\psi=0 ,
\tag 2.4$$
then the elements of Bloch's wave basis
$\{\Psi_{i}(x);i=1$,$\ldots$,$n\}$ of (2.4)
is the
components of the vector $\vec{\Psi}$ associated with the vector
$n$-dimensional representation of $WGL(n)(WSL(n))$ [6]:
 $$
\Psi_{i}(x)=\prod_{s=1}^{i-1} \lambda_{s}{(\lambda_{i}-\lambda_{s})}^{-1}
\exp(\vec{\omega_{1}}\vec{\phi}(x)) \oint_{x(\sigma)}^{x(\sigma+2\pi)}
{d}\eta_{1}
\ldots \oint_{\eta_{i-2} (\sigma)}^{\eta_{i-2} (\sigma+2\pi)}
d \eta_{i-1}
$$
$$
\exp(-\vec{\alpha}_{1}
\vec{\phi}(\eta_{1})) \ldots
\exp(-\vec{\alpha}_{i-1}\vec{\phi}(\eta_{i-1})) \quad .
\tag 2.5$$
Functions $\Psi_i(x)$ satisfy the periodicity condition
$\Psi_i(\sigma_0+2\pi)=\lambda_{i}\Psi_i(\sigma_0)$ with
$\lambda_{i}=\exp(\vec{p} \vec{h}_{i})$ and choise of the
initial point $x(\sigma_0)$ is unimportant.

In the case of $GL(n)$ group vectors $\vec{h_{i}}$ constitute
the orthonormal basis in the $n$-dimensional vector space:
$<\vec{h}_{i},\vec{h}_{j}>={\delta}_{ij}$, $\vec{\phi}(z)
\vec{h_{i}}={\phi}_{i} (z) $.

Let us remind the pseudodifferential operator definition:
$$
[\partial^{i},\partial^{j}]=0,\quad [\partial^{i},a(x)]=\sum
^{\infty}_{p=1}{{i(i-1)\ldots(i-p+1)} \over {p!}}
a^{(p)}(x)\partial^{i-p},
\quad i,j \in{Z}.
\tag 2.6$$
Given a pseudodifferential operator $P=\sum p_{i}\partial^{i}$
we define its non-commutative residius and Adler's trace ([14]) as:
$$
res(\sum p_{i}\partial^{i})=p_{-1},\quad Tr(\sum
p_{i}\partial^{i})=
\oint {{dx}\over{2{\pi}i}}res(\sum p_{i} \partial^{i}).
$$
Taking into account eqs.(2.2) and (2.3) it is
not difficult to show that the Poisson bracket of the hamiltonian
$$
\eqalign{H&=Tr(YL)=\sum^{n-1}_{i=0}
\oint{{dx}\over{2{\pi}i}}a_{i}(x)
W_{n-i}(x) ,\cr
Y&=\sum^{n-1}_{i=0}\partial^{-i-1}a_{i}\cr}$$
with $\Psi$s
(i.e. $W$-algebra action on $\Psi$s) is
given by
$$\delta_{H}\Psi_{j}=\{l_{Y},\Psi_{j}\}=
-{(YL)}_{+}\Psi_{j} \quad .
\tag 2.7$$
As usual, ${(YL)}_{+}$ stands for the differential
part of the pseudodifferential operator which is polynomial
in $\partial$ (including free term).
Really, it is enough to prove(2.7) for $\Psi=\exp(\phi_1)$
and the transformation properties of other fields will
be the same.
Denoting
$$
L{(\partial-\phi_1')}^{-1}=(\partial-\phi_n')...
(\partial-\phi_2')
\tag 2.8$$
and using that
$$
{(\partial-\phi_1')}^{-1}=\exp({\phi_1}){\partial}^{-1}\exp({-\phi_1})
\tag 2.9$$
one can immeadetely obtain:
$$\eqalign{&\delta_{H}(\exp{\phi_1})(z)=\oint{{d\zeta}\over{2{\pi}i}}
\{res[Y(\zeta)L(\zeta)],\exp{\phi_1}(z)\}=\cr
&=-\oint{{d\zeta}\over{2{\pi}i}}res[Y(\zeta)L(\zeta)
{(\partial-\phi_1'(\zeta))}^{-1}\delta(\zeta-z)]\exp(\phi_1(z))\}=\cr
&=-\oint{{d\zeta}\over{2{\pi}i}}res[Y(\zeta)L(\zeta)\exp(\phi_1(\zeta))
{\partial}^{-1}\exp(-\phi_1(\zeta))\delta(\zeta-z)]\exp(\phi_1(z))\}=\cr
&=-res[Y(z)L(z)\exp(\phi_1(z)){{\partial}_z}^{-1}]
=-{(YL)}_{+}\exp(\phi_1(z))\quad . Q.E.D.\cr}
\tag 2.10$$

\vglue 12pt
\line{\tenbf 3.QUANTUM CASE \hfil}
\vglue 5pt
\line{\tenbf 3.1 Basic Definitions. ([6]). \hfil}
In the quantum case we shall assume that all operators
are defined on $\{ C \setminus 0 \}$.
Let $T$ be the notation of the
operator product $T$:
$$
T(A(\zeta)B(z))=\quad \{\hbox{singular terms }\}\quad + :A(z)B(z):\quad .
\tag 3.1$$
As is well-known, singular terms under $z\rightarrow {\zeta}$
determine commutation relation
for generators of the fields [4].
The Poisson brackets (2.2) for the Darboux's variables
$\phi^{a}$($a=1,\ldots,n$)
(2.1) correspond to the operator product of the free fields:
$$
T(\phi^{a}(\zeta)\phi^{b}(z))= {{\delta^{ab}}\over k} \log(\zeta-z)
+O(\zeta-z) ,
\tag 3.2$$
or, equivalently:
$$[a_{m}^{b},a_{l}^{c}]={l\over k}\delta_{m+l,0}\delta^{bc},\quad
[p^a,q^b]={{2{\pi}i} \over k} \delta^{a,b}$$
$$
[p^a,a_m^b]=[q^a,a_m^b]=0.
\tag 3.3$$
We can consider $k$ as the deformation parameter.
Then $W$-currents are defined via quantum Miura transformation [5]:
$$
{(\alpha_0 \partial)}^{n}+\sum W_{i}(z){(\alpha_{0}
\partial)}^{n-i}  =\quad
:\prod_1^n(\alpha_0\partial-\vec{h}_{i}\vec{\phi}'(z)):,
\tag 3.4$$
where $\alpha_0$ is some constant.
It was proved in ref.[5] that quantum operators in the left hand side
of (3.4) form the associative quadratic algebra.
Demanding that vertex operators
$\exp(\vec{\alpha_{i}}\vec{\phi})$ has conformal dimension
$\Delta=1$ in the $SL(n)$-theory yields [4]:
$$\alpha_0=1-{1\over k}.
\tag 3.5$$

Now, let
us consider the n-dimensional completely degenerated highest weight
representation of $W$ algebra determined by
highest weight operator
$$F(\vec{\omega}_{1}\|z)=
:\exp(\vec{\omega_{1}}\vec{\phi}): \quad .
$$
The representation space is covered by the following
$n$ linearly independent operators :
$$
\eqalign{F_{i}(\vec{\omega_1}\|z)=&\prod_{s=1}^{i-1}{{\lambda_{i}}
\over{\lambda_{s}-\lambda_{i}}}\oint_{C_1}
{{d\eta_{1}}\over {2{\pi}i}}\ldots\oint_{C_{i-1}}
{{d\eta_{i-1}}\over{2{\pi}i}}
T[:\exp(-\vec{\alpha}_{1} \vec{\phi}(\eta_{1})): ...\cr
&\ldots:\exp(-\vec{\alpha}_{i-1}\vec{\phi}(\eta_{i-1})):
:\exp(\vec{\omega}_{1}\vec{\phi}(z))]: \quad .\cr}
\tag 3.6$$
The integration contours in this formula
are choosen in accordance
with the standard Felder's prescription (surrounding all
singular points [22]).
\vglue 5pt
\line{\tenbf 3.2 W-algebra action. \hfil}
\vglue 5pt
Let us define the action of $W$-currents on local fields
of the theory as [4]:
$$
\delta_{H_Y} S=k\oint_{C} {dz \over {2{\pi}i}}
\sum_{l=1}^{n}
a_{n-l}(z)T(W_{l}(z)S),
\tag 3.7$$
where Hamiltonian $H_Y$ has the form:
$$H_Y=\oint_{C}{dz \over {2{\pi}i}}\sum_{l=1}^n
a_{n-l}(z)W_{l}(z)=Tr(YL) \quad,
$$
$$
Y=\sum^{n-1}_{i=0}\partial^{-i-1}a_{i} \quad ,
$$
and integration contour $C$ surrounds all
singular points of integrand.
The action of nilpotent part of quantum group $U_q(gl(n))$ ($U_q(sl(n))$)
represented by screening factor has to
be commutative with the action of W-algebra (this
property can be considered as the definition of $W$ algebra [11,12]). Indeed,
it was proved in ref.[5] that
$$
T(W_{i}(z)\exp(-\vec{\alpha}_{i}\vec{\phi}(\zeta))={\partial\over {
\partial\zeta}}X_{i}(z,\zeta) +o(1) \quad ,
$$
where $X_{i}$  is some local operator. Therefore
we can afford to ignore screening operators which do not
change the transformation properties of the fields.
In this section we will consider only the highest weight
operator  which for $WGL$ has the form
$F_{1}(\vec{\omega_1}\|z)=F_1(z)=:\exp(\phi_1(z)):$.
But all formulae will be valid for any invariant operator
$F_i$

In analogy with classical case we will first evaluate the
$W$-algebra action on the operator $F_1$.
Using quantum Miura transformation, one can express
$W_{i}$ as normal ordered differential polynomials in terms of
the fre fields $\phi_{a}$ and then perform Vick's
pairing with $F_{1}(z)$ (only one in $GL(n)$ case).
Plugging (3.4) into (3.7) after some simple manipulations
one finds:
$$
\eqalign{\delta_{H}F_1(z)&=k\oint {{d\zeta}\over {2{\pi}i} }
  a_{n-i}(\zeta)T(W_{i}(\zeta)F_1(z))\cr
&=:\exp(\phi_{1}-{{\phi_{1}}\over {\alpha_{0}}}){(\tilde{Y} \tilde{L})}
  _{+}\exp({{\phi_{1}}\over {\alpha_{0}}}): \quad ,\cr}
\tag 3.8$$
where the notation
$$
\tilde{Y}=\sum{(\alpha_{0}\partial)}^{-i-1}a_{i},
$$
$$
\tilde{L}=(\alpha_0\partial)^n+\sum W_{i}^\phi{(\alpha_{0}\partial) }^{n-i}=
:\prod_1^n(\alpha_0\partial-\phi'_i):
$$
is introduced.
Note that expressions like $\exp(\phi_1
(1-{1\over{\alpha_{0}}}))(W^\phi)^{(q)}{(\alpha_0\partial)}^{p}
\exp({{\phi_1}\over
{\alpha_0}})$ are rather formal. We need to provide
all differrentiations and then turn to the
form $:Pol(\phi)F_1:$, where symbol $Pol$ stands for
some differential polinomial in $\phi$.
For example, term $\exp(\phi_1
(1-{{1}\over {\alpha_{0}}})) {(\alpha_0\partial)}^2
\exp({{\phi_1}\over{\alpha_{0}}})$
has to be understood as
$[(\alpha_{0}{\partial}^{2}{\phi}_1
+{(\partial {\phi}_1) }^{2})F_1]$
and so on.
It is significant that the notation ": :" in (3.8) implies
the ordering of
$\phi$-fields only. However this is not yet the whole
story . To provide the ordering of $W_{i}$-fields
we must to change polynomials of $\phi$'s in (3.8)
(which formally correspond to $W$-currents) by the first
non-vanishing terms of its operator product with
$F_1=:\exp({\phi}_{1}):$. Let us denote
coefficient under ${W_{(q)}}^{(l)}{\partial}^{p}$
in the formula (2.7) as
$C_{q}^{l,p}$ . Clearly, that $C_{q}^{l,p}$ after such
corrections will change on some value $\delta C_{q}^{l,p}$.
The difficulty in determination of these quantum corrections
is that non-vanishing terms in
T($({W_{q}}^{\phi})^{(l)}{F_1}^{(p)}$) include formal differential
polynomials in the $\phi$ of the form
$({W_{q}}^{\phi})^{(l-1)}{\partial}^{p},\quad
({W_{q}}^{\phi})^{(l)}{\partial}^{p-1},\quad
({W_{q-1}}^{\phi})^{(l)}{\partial}^{p},\quad \ldots$.
The laters need to be ordered too and so on.
\vglue 5pt
\line{\tenbf 3.3 Examples.\hfil}
\vglue 5pt
To llustrate the general
procedure of ordering,
let us now consider examples of $W_2$, $W_3$ and $W_4$
-currents action on $F_1$.
At first we express $W_1, W_2, W_3$-currents through the free fields
using quantum Miura transformation:
$$
\eqalign {
W_1&=-\sum_{i=1}^n\phi'_i \quad ,\cr
W_2^{\phi}&=\sum_{i<j}^n \phi'_j \phi'_i-\sum_{i}^n (n-i)
\alpha_0 \phi''_i \quad ,\cr
W_3^{\phi}&=-\sum_{i<j<k}^n \phi'_k \phi'_j \phi'_i+
\sum_{i<j}^n [(n-j)\alpha_0 \phi''_j\phi'_i+\cr
&+(n-i-1)\phi'_j\phi''_i ]-\sum_{i}^n {{(n-i)(n-i-1)}\over 2}
{\alpha_0}^2 \phi'''_i \quad ,\cr}
\tag 3.9$$
{}From this representation one can immideately find
transformation of highest weight operator under the action of $W_2$-current:
$$
T(W_2(\zeta)F_1(z))=\sum_{i>1}^n [{1 \over k(\zeta-z)}\phi'_i(\zeta)-
\partial^2_{\zeta} {1 \over k(\zeta-z)}\alpha_0(n-i)] \quad ,
\tag 3.10$$
$$
\eqalign{
\delta_{H_2}F_1&=k\oint {{d\zeta}\over{2{\pi}i}}a_{n-2}(\zeta)
T(W_{2}(\zeta):\exp(\phi_1(z)):=\cr
&=-:[\alpha_0(1-n)a'_{n-2}+a_{n-2}(W_1+\partial)]F_1: \quad.
\cr}
\tag 3.11$$
We have no need for additional ordering procedure because
the expression $:W_1F_1:$ coinsides with $:W_1^{\phi}F_1:$.
First non-trivial example in which
the ordering problem arises is the
$W_3$-current action. As mentioned above, the
normal ordering after Vick's pairing implies
the ordering of the free fields only. While $W_2^\phi$-current
which appeared in the expansion
have to be ordered under the scheme:
$$
:W_2^{\phi}(z)F_1(z): \rightarrow
\oint{{d\zeta}\over {2{\pi}i}} {T(W_2(\zeta)F_1(z)) \over (\zeta-z)}
$$
It follows from the operator expansion (3.10) that:
$$
:W_2F_1: =
\oint {{d\zeta}\over{2{\pi}i}} {
T(W_{2}(\zeta)F_1(z)\over (\zeta-z)}=
:({W_2}^{\phi}-{1\over k}W_1'-{1\over k}\phi_1'')F_1:.
\tag 3.12  $$
By definition, the action of $W_3$ current on the reguralized exponent has the
form:
$$
\delta_{H_3}[F_1(z)]=
k\oint {{d\zeta}\over{2{\pi}i}}a_{n-3}(\zeta)
T(W_{3}(\zeta):\exp(\phi_1(z)): \quad ,
\tag 3.13a$$
where T-product of $W_3$ with $F_1$ after Vick's pairing
and expansion under the powers of $(\zeta-z)$ is given by:
$$
\eqalign{
&T(W_3(\zeta)F_1(z))=
:{1 \over k}[-{1\over{(\zeta-z)^3}}(n-2)(n-1){\alpha_0}^2+\cr
&+{1\over{(\zeta-z)^2}}(n-2){\alpha_0}(W_1(z)+\phi_1'(z))+
{1\over{(\zeta-z)}}[-(W_2^{\phi}(z)-\cr
&-{1\over k}W_1'(z)-{1\over k}\phi_1''(z))
-W_1(z)\phi_1'(z)-(\phi_1''(z)+{\phi_1'}^2(z))+\cr
&+((n-2){\alpha_0}-{1\over k})W_1'(z)]F_1(z):+
\{\hbox{regular terms }\} \quad .\cr}
\tag 3.13b$$
Now substituting eqs.(3.12),(3.14) into (3.13b) we obtain after
the contour integration:
$$
\eqalign{
&\delta_{H_3}[F_1(z)]= :[-{{(n-2)(n-1)}\over{2}}
{\alpha_0}^2a''_{n-3}(z)+
(n-2){\alpha_0}a'_{n-3}(z) [W_1(z)+\partial_z]+\cr
&a_{n-3}(z)[-W_2(z)
-W_1(z)\partial_z-\partial_z^2+((n-2){\alpha_0}-
{1\over k})W_1'(z)]F_1(z): .\cr
}
\tag 3.14
$$
Analogous but slightly tedious computation for $W_4$-current
yields:
$$
\eqalign{
&\delta_{H_4}[F_1(z)]=
\oint {dz \over {2{\pi}i}}a_{n-4}(\zeta)
T(W_{4}(\zeta):\exp(\phi_1(z)):)=\cr
&=:(-{{(n-3)(n-2)(n-1)}\over{6}}{\alpha_0}^3a_{n-4}'''-
{{(n-3)(n-2)}\over 2}{\alpha_0}^2a_{n-4}''[(W_1(z)+\partial)\cr
&+(n-3)\alpha_0 a'_{n-3}[W_2+W_1\partial+{\partial}^2
-((n-2)\alpha_0-{1\over k})W_1']\cr
&-a_{n-4}[W_3+W_2{\partial}+W_1{\partial}^2+\partial^3-
((n-3)\alpha_0-{1\over k})W_2'
-((n-3)\alpha_0-{2\over k})W_1'\partial+\cr
&{1\over 2}(((n-3)\alpha_0-{1\over k})((n-2)\alpha_0-
{1\over k})+{1\over k}]W_1'')]
F_1(z): \quad ,\cr}
\tag 3.15$$
where formulae for the following normal ordering
expressions are used:
$$\eqalign {
&:W'_2(z)F_1(z)):=\oint {{d\zeta}\over{2{\pi}i}}
{T(W'_{2}(\zeta):\exp(\phi_1(z)):\over (\zeta-z)}=\cr
&=:[{W'_2}^{\phi}-
{1\over 2k}(W''_1+\phi'''_1)]F_1: \quad ,\cr
&:W_2(z)F'_1(z)):=\oint {{d\zeta}\over{2{\pi}i}}
{T(W_{2}(\zeta):\phi_1(z)\exp(\phi_1(z)):\over (\zeta-z)}=\cr
&=:[{W_2}^{\phi}\partial-{1\over {2k}}
(2W'_1\phi''_1+\phi'''_1+2\phi'_1\phi''_1+W''_1)]F_1(z) \quad ,\cr
&:W_3(z)F_1(z):=\oint {d\zeta \over 2{\pi}i}
{T(W_3(\zeta):\exp(\phi_1(z)): \over (\zeta-z)}=\cr
&=:(W_3^{\phi}-{1\over k}({W'_2}^{\phi}-{1\over 2k}(W''_1+\phi'''_1)+\cr
&+(-{(n-2)\alpha_0\over 2}+
{1\over {2k}})W''_1+W'_1\phi'_1+
W_1\phi''_1+\cr
&+2\phi'_1\phi''_1+
({{n\alpha_0+{1\over k}}\over 2}\alpha_0\phi'''_1)F_1(z): \quad .\cr}
\tag 3.16 $$
One can see, for example, that quantum correction $\delta C_{1}^{0,3}$
depends not only from the ordering of
$W_2^{\phi}F'_1$,
${W'}_2^{\phi}F_1$ and
$W_3^{\phi}F_1$
terms appeared after Vick's pairing but also
from the ordering of
${W'}_2^{\phi}$-current in the eq.(3.15c).
It is clear that for large $n$ direct procedure of finding
$\delta C_{q}^{l,p}$ becomes more involved.

To find quantum $L$-operator one can determine quantum corrections
to the expressions of the form ${W_l}^\phi F^{(n-l)}_1$. Explicit
calculations implies that for small n the sum of corrections to
the expression
$
:[{(\alpha_0 \partial)}^{n}+\sum W_{i}^\phi(z){(\alpha_{0}
\partial)}^{n-i}]F_1:$
has the form
$:\sum_{p+q+l=n,l>0} \delta C_{q}^{l,p}{\partial}^{l}W_{q}F^{(p)}:$,
where $\delta C_{q}^{l,p}$ have precisely the same value as before.
Let us give exact formulae for small n:
$$
\eqalign{
n=2  &:[W_2+W_1\partial+{1\over k}W'_1
+{\partial}^2]]F_1: =0,\cr
n=3  &:[W_3+{1\over k}W'_2+{1\over {k^2}}W''_1+
(W_2+{2\over k}W'_1)\partial+\cr
&+W_1{\partial}^2+{\partial}^3]F_1: =0,\cr
n=4  &:[W_4+{1\over k}W'_3+{1\over {k^2}}W''_2+{1\over {k^3}}W'''_1+\cr
&+(W_3+{2\over k}W'_2+{1\over {k^2}}W''_1)\partial+\cr
&+(W_2+{3\over {k}}W'_1){\partial}^2+{\partial}^3]F_1:=0 .\cr}
\tag 3.17$$
Here we replace expressions for  $:W_3F_1:$, $:W_2F_1:$, $:W'_2F_1:$ and
$:W_2F'_1:$
using eqs.(3.12) and (3.16). Unfortunately,
formulae for
$:W_4F_1:$, $:W_3F_1:$, $:W'_3F_1:$ etc. in the terms
of free fields are too cambersome to be represented here.
\vglue 12pt
\line{\tenbf 4. QUANTUM DEFORMATION OF OPERATORS. \hfil}
\vglue 5pt
\line{\tenbf 4.1 Right derivations. \hfil}
\vglue 5pt
Let us now introduce the convinient notation of the right derivation
$\bar{\partial}$ and define the deformation of the classical
operators $Y$ and $L$ preserving
Adler's trace of the product $(YL)$.
$$
f(x)\bar{\partial}=f'(x)+\bar{\partial}f(x),
\tag 4.1$$
where right derivations commutes with ordinary ones
$[\bar{\partial},\partial]=[\bar{\partial},{\partial}^{-1}]=0$.
Define integral simbol ${\bar{\partial}}^{-1}$ as:
$$\bar{\partial}^{-1}\bar{\partial}=\bar{\partial}\bar{\partial}^{-1}=1,
[{\bar{\partial}}^{-1},\partial]=[{\bar{\partial}}^{-1},{\partial}^{-1}]=0,
$$
$$[f(x),\bar{\partial^{i}}]=\sum
^{\infty}_{p=1}{{i(i-1)\ldots(i-p+1)} \over {p!}}
{\bar{\partial}}^{i-p}f^{(p)}(x),
\quad i,j \in{Z}.
$$
$$
(\partial+{{\bar{\partial}}\over{k}})^{-l}={\partial}^{-l}(1+
{{{\partial}^{-1}\bar{\partial}}\over k})^{-l}={\partial}^{-l}-
{{l\over k} {\partial}^{-l-1}\bar{\partial}}+
{{l(l-1)}\over {2k^2}}{{\partial}^{-l-2}{\bar{\partial}}^2}+...
\tag 4.2$$
Right Adler's trace of the pseudodifferential simbol
$P=\sum{\bar{\partial}}^{i}f_{i}(x)$ like an ordinary one
is given by
$$
\bar{res}P=\bar{res}\sum{\bar{\partial}}^{i}f_{i}(x)=f_{-1}.
    $$
So we have two copies of derivatives. Such as left and
right symbols commutes to one another,
there are no additional difficulties with
definition of Adler's trace for bi-pseudodifferential
operators of the form $\sum_{i,j}\bar{\partial}^if_{i,j}(x)\partial^j$:
$$
res[\sum_{i,j}\bar{\partial}^if_{i,j}(x)\partial^j]=
\sum_{i}\bar{\partial}^if_{i,-1}(x)  \quad,$$
and
$$
\bar{res}[\sum_{i,j}\bar{\partial}^if_{i,j}(x)\partial^j]=
\sum_{j}f_{-1,j}(x)\partial^j
\quad .$$
\vglue 5pt
\line{\tenbf 4.2  Deformed $Y$ and $L$ operators.\hfil}
\vglue 5pt
In this section we propose some deformation of $Y$ and
$L$ operators. Our consideration is based on the simple
\vglue 5pt
\noindent
{\tenbf Lemma.} Quantum Miura transformation can be rewriting as
the equality of two bi-differential operators
$$
\bar{L}=( \partial+{{\bar{\partial}}\over{k}})^{n}+\sum W_{i}(z)
(\partial+{{\bar{\partial}}\over{k}})^{n-i}  =\quad
:\prod_1^n(\partial+{{\bar{\partial}}\over{k}}-{\phi}_{i}(z)):   .
\tag 4.3$$
{\tenbf Proof.} W-currents expressed from this formula
via free bosonic fields
coincide with an ordinary ones due to the following
property of derivations:
$$(\partial+{{\bar{\partial}}\over {k}})f(x)=(1-{1\over {k}})f'(x)
+f(x)(\partial+{{\bar{\partial}}\over {k}}) \quad Q.E.D.$$
It would be reasonable to suggest that $L$ operator
can be constructed from $\bar{L}$. Let us introduce quantum $L$-operator
$\hat{L}$ as
$$
\eqalign{
\hat{L}&=1(\bar{L})=\bar{res}[\bar{\partial}^{-1}
\bar{L}]\cr
&={\partial}^n+W_{1}{\partial}^{n-1}+ [W_{2}+{{n-1}\over {k}}{W_1}')
{\partial}^{n-2}+\cr
&+[W_{3}+{{n-2}\over {k}}{W_2}'
+{{(n-2)(n-1)}\over {2k^2}}{W_1}'']{\partial}^{n-3}+... \quad .\cr}
\tag 4.4$$
For for $n=2,3,4$ one can recognize in this formula eqs.(3.17).
We will show further that eq.(4.4) is the proper expression for the
quantum $L$-operator in case of an arbitrary natural $n$.

We define quantum deformation for
the pseudodifferential symbol $Y$ as dual to the $L$
with respect to the bilinear form given by Adler's trace:
$$
\eqalign{
\hat{Y}&=\bar{res}[\bar{Y}\bar{\partial}^{-1}]=
\bar{res}\sum[(\partial+{{\bar{\partial}}\over{k}})^{-l-1}
a_l{\bar{\partial}}^{-1}]\cr
&=[{\partial}^{-l-1}a_l+{{l+1}\over k} {\partial}^{-l-2}a_l'+
{{(l+1)l}\over {2k^2}} {\partial}^{-l-3}a_l''+... \quad .\cr}
\tag 4.5$$
It is easy to prove from the properties of
bi-pseudodifferential operators the following usefull
\vglue 5pt
\noindent
{\tenbf Lemma.}

\noindent
(i)
$$ res (YL)=res (\hat{Y}\hat{L}) \quad .
$$
(ii) Hamiltonian $H$ is given by
$$
H=Tr(YL)=Tr((\hat{Y}\hat{L}))=k\oint_{C}
{dz \over {2{\pi}i}}\sum_{l=1}^{n}a_{n-l}(z)W_{l}(z) \quad.
\tag 4.6$$
(iii)
$$(\hat{Y}\hat{L})=\bar{res}[\bar{Y}\bar{\partial}^{-1}\bar{L}] \quad .
\tag 4.7$$

\vglue 12pt
\line{\tenbf 5. QUANTUM FORMULAE. \hfil}
\vglue 5pt
\line{\tenbf 5.1 Quantum $L$ operator. \hfil}
\vglue 5pt
At first we will prove that quantum analogue of the $L$-operator
is given by $\hat{L}$:

\noindent
{\tenbf Proposition.}
$$
:[\hat{L}F_s]:=0 \quad s=1,...,n \quad .
\tag 5.1$$
\noindent
{\tenbf Proof.} It is convinient for us to extract derivations in $\hat{L}$:
$$\hat{L}=\sum_i res[\hat{L}\partial^{-i-1}]\partial^{i} \quad .$$
Now using this formula and the definition of the normal
ordering one can easily obtain:
$$
\eqalign{
&:\hat{L}F_s(z):=:\sum_i res(\hat{L}\partial^{-i-1})
F^{(l)}_s(z)):=\cr
&=:\sum_i[\oint{d\zeta \over 2{\pi}i(\zeta-z)}
T(\bar{res}[res(\bar{\partial}^{-1}_{\zeta}\bar{L}_\zeta^\phi
\partial^{-i-1}_{\zeta})]F^{(i)}_s(z)]:=\cr
&=:\sum_i[\oint{d\zeta \over 2{\pi}i(\zeta-z)}
\bar{res}[res(\bar{\partial}^{-1}\bar{L}_\zeta^\phi
-\bar{\partial}^{-1}\bar{L}_\zeta^\phi
(\partial_\zeta+{1 \over k}\bar{\partial}_\zeta-\phi'_1
(\zeta))^{-1}\cr
&\partial_z^i({F_s(z)\over (\zeta-z)})\partial^{-i-1}_\zeta)]
F_s^{(i)}(z):=
:[\hat{L}^{\phi}F_s(z)]:-\cr
&-:\sum_{il}\{\oint{{d\zeta}\over {2{\pi}i(\zeta-z)}}\bar{res}[(-\bar
{\partial}^{-1}\bar{L}_\zeta^\phi
(\partial_\zeta+{1 \over k}\bar{\partial}_\zeta-\cr
&-\phi'_1(\zeta))^{-1}]{C^i_l l!\over k(\zeta-z)^{l+1}}
\partial^i_\zeta\}
F^{(i-l)}_s(z): \quad .\cr}
$$
Let us introduce the following shorthand notation
$$
-\bar{res}[{\partial}^{-1}\bar{L}^\phi(\partial+{1 \over k}
\bar{\partial}-
\phi'_1)^{-1}]=-\sum^{n-1}_{j=0}\bar{res}[\bar{Q}_j\partial^j],
\tag 5.2$$
where $\bar{Q}_j$ stands for right pseudodifferential
operator $\bar{Q}_j=\sum_i\bar{\partial}^iq_{ij}$ and $q_{ij}$ are
some differential polynomials in terms of the $\phi$'s.

After the differentiation
procedure we obtain taking the residus:
$$
\eqalign{
&:[\hat{L}F_s(z)]:=\cr
&=:[\hat{L}^{\phi} F_s(z)]:+
:\sum_{ijlp}[\oint{{d\zeta}\over {2{\pi}i}}
\bar{Q}_j{l!\over (\zeta-z)^{l+p+1}} C^j_pC^i_l
{\partial}_{\zeta}^{j-p-i-1}]F_s^{(i-l)}(z):=\cr
&=:[\hat{L}^{\phi}F_s(z)]:
-:\sum_{ijl}{\bar{Q}_i\bar{\partial}^{(l+j-i+1)} \over k(l+j-i+1)}
(-1)^{j-i+1}
{j!\over (j-i)!l!(i-l)!}F_s^{(i-l)}(z): \quad .\cr
}
$$
It is rather simple calculation to show that sum in the second term
reduces to
$$[-\sum_j(\bar{Q}_j\bar{\partial}\partial^j/k)\exp(\phi_1)]\quad .
$$
In the other hand, from the structure of the first  term we find
$$
\eqalign{
&:\bar{res}[\bar{\partial}^{-1}\bar{L}^\phi]F_s=\bar{res}
[\bar{\partial}^{-1}
(\partial+{1\over k}\bar{\partial}-\phi'_n)
...(\partial+{1\over k}\bar{\partial}-\phi'_2)\cr
&(\partial+{1\over k}\bar{\partial}-\phi_1)]F_s:=
:\bar{res}[\bar{\partial}^{-1}\bar{L}^\phi
(\partial+{1\over k}\bar{\partial}-\phi_1)^{-1}
{1\over k}\bar{\partial}]F_s:=\cr
&=:\sum_j(\bar{Q}_j\bar{\partial}\partial^j/k)F_s: \quad .\cr}
\tag 5.3$$
Finally we get
$$
:[\hat{L}F_s]:=0 \quad .$$
Q.E.D.
\vglue 5pt
\line{\tenbf 5.2 W-algebra action. Proof. \hfil}
\vglue 5pt
\noindent
Now turn our attention to the
finding of the $W$ algebra action.
Following the scheme above we will prove that
there exist simple quantum deformation of
formulae (2.7).
\vglue 5pt
\noindent
{\tenbf Proposition.}
$$\delta_{H} F=-:[{(\hat{Y}\hat{L})}_{+} F_s]: \quad s=1,...,n \quad,
\tag 5.4$$
where $\hat{Y}$ and $\hat{L}$ are given by eqs. (4.4) and (4.5).

\noindent
{\tenbf Proof.} Explicit calcucation shows that eqs.(3.11),(3.14),(3.15)
can be recast in such a form.
To prove this in general case let us represent both expressions
in (5.4) through the free fields and show that
they are identical.

Unique Vick's pairing (due to eq.(3.2)) of $H$ and $F_s$ can
be found as for the classical case :
$$
\eqalign{
&\delta_{H}(F_s(z))=kT(Tr[YL]F_s(z))=\cr
&=kT(Tr[\bar{res}(\bar{Y}\bar{\partial}^{-1}\bar{L})]F_s(z))))=\cr
&=-:(\oint {{d\zeta}\over{2{\pi}i}}
\bar{res}[res(\bar{Y}_\zeta{\bar{\partial}_{\zeta}}^{-1}\bar{L}^\phi_\zeta
(\partial_{\zeta}+{1\over k}\bar{\partial}_{\zeta}-\phi'_1(\zeta))^{-1}
{1\over {\zeta-z}})]
F_s(z)):=\cr
&=-:\bar{res}[res(\bar{Y}(z){\bar{\partial}_{z}}^{-1}\bar{L}^\phi_z
(\partial_{z}+{1\over k}\bar{\partial}_{z}-\phi'_1(z))^{-1})]
F_s(z): \quad .\cr}
\tag 5.5$$
This is exactly the same equation as  (3.8).
Let us denote for the sake of simplicity
$$[\bar{Y}\bar{\partial}^{-1}\bar{L}(\partial+{1\over k}
\bar{\partial}-\phi'_1)^{-1}]=\sum_j\bar{P}_j\partial^j\quad ,
\tag 5.6$$
where symbol $\bar{P}_j$ stands for  some right differential operator
$\bar{P}_j=\sum_i \bar{\partial}^if_{ij}$
with differential polinomial of $\phi$ as coefficients.
Now one can rewrite the formula (5.5) as:
$$\eqalign {
\delta_{H}(F_s(z))&=-:\sum_j\bar{res}[res(\bar{P}_j\partial^j)]
F_s(z): =\cr
&=-:\bar{res}[\bar{P}_{-1})]F_s(z): \quad .\cr}
\tag 5.7  $$
Consider the expression $:[-{(\hat{Y}\hat{L})}_{+}F_s]:$.
At first let us note that
$$(\hat{Y}\hat{L})_{+}=\sum_{i\geq 0}(\hat{Y}\hat{L})_{i}\partial^i
=\sum res(\hat{Y}\hat{L}\partial^{-i-1})\partial^i   .
$$
Try to find now the operator product
$T([(\hat{Y}\hat{L})_{+}]_{i}(\zeta) F^{(i)}_s(z))$.
Of course,
after Vick's pairing and expansion under the powers of $(\zeta-z)$ in the point
$z$
we don't obtain quantum corrections.
However the special structure of $(\hat{Y}\hat{L})_{+}=
\bar{res}(\bar{Y}\bar{\partial}^{-1}\bar{L})_{+}$ make it possible
to apply some trick and find the proper expression.
$$
\eqalign {
&\bar{res}[
:\bar{Y}_z\bar{\partial}^{-1}_z\bar{L}_z)_{+}F_s(z):]\sim
\sum_i\bar{res}[
T[(\bar{Y_\zeta}\bar{\partial}^{-1}_\zeta\bar{L}_\zeta)_{+}]_i
F^{(i)}_s(z)]=\cr
&=\sum_i
T \{ \bar{res}[res(\bar{Y}_\zeta\bar{\partial}^{-1}_\zeta
\bar{L}_\zeta\partial^{-i-1}_\zeta)]
F^{(i)}_s(z) \}=\cr
&=:(\hat{Y}_z{\hat{L}^{\phi}_z)}_{+}F_s(z):-
:\sum_{iq}\bar{res}[res(\bar{Y}^\phi_{\zeta}
\bar{\partial}^{-1}_{\zeta}\bar{L}^\phi_{\zeta}
(\partial_{\zeta}+{1\over k}\bar{\partial}_{\zeta}
-\phi_1)^{-1}(\zeta) \cr
&[\partial_z^q{1\over{k(\zeta-z)}}]\partial_{\zeta}^{-i-1})]
C^i_qF_s(z)^{(i-q)}: \quad .\cr}
\tag 5.8$$
Let us consider the first term in this formula:
$$\eqalign{
&:(\hat{Y}\hat{L}^{\phi})_{+}F_s):=
:(\bar{res}[(\bar{Y}\bar{\partial}^{-1}\bar{L}])_{+}]F_s:=\cr
&=:(\bar{res}[\bar{Y}\bar{\partial}^{-1}\bar{L}^\phi(\partial+
{1\over k}\bar{\partial}-\phi'_1)^{-1}
(\partial+{1\over k}\bar{\partial}-\phi'_1)])_{+}F_s:=\cr
&=:\sum_{j\geq 0}\bar{res}[\bar{P}_j\partial^j
(\partial+{1\over k}\bar{\partial}-\phi'_1)]_{+}
F_s:=\cr
&=:\bar{res}[{1\over k}\sum_{j \geq 0}\bar{P}_j\bar{\partial}\partial^j+
\sum_{j \geq 0}\bar{P}_j\partial^{j+1}+\cr
&+\bar{P}_{-1}-\sum_{j\geq 0} \bar{P}_j\partial^j\phi'_1]F_s:
=:\bar{res}
[{1\over k}\sum_{j \geq 0}\bar{P}_j\bar{\partial}\partial^j+\bar{P}_{-1}]
F_s: \quad .\cr
}
\tag 5.9$$
Direct calculation shows that quantum corrections in the
formulae (5.8) cancel sum in the
expression (5.9) and the remainder term $:\bar{res}\bar{P}_{-1}F_s:$ gives
formula (5.7). Indeed let us turn to the
second term in (5.8).
After the changing (5.6) and differentiation one can immediately
obtain that singular terms in (5.8) can be represented as:
$$
\eqalign {
&-\sum_{iq}{1\over k}\bar{res}[res(\bar{P}_j(\zeta)\partial^j_\zeta
(\partial_z^q{1\over{(\zeta-z)}})\partial_{\zeta}^{-i-1})]
C^i_qF_s^{(i-q)}(z)=\cr
&=-{1\over k}\sum_{ijql}\bar{res}[
C^j_lC^i_q\bar{P}_j(\zeta)
({(l+q)!(-1)^l\over{(\zeta-z)}^{l+q+1}}\delta_{j-l-i-1,-1})
F_s^{(i-q)}(z) \quad .\cr}
\tag 5.10$$
We have need for one more
\vglue 5pt
\noindent
{\tenbf Lemma.} Expression (5.10) gives proper
formulae for quantum corrections
if the expansion of
$\bar{res}[\bar{P}_j]$ in terms of $(\zeta-z)$
has the form:
$$
\bar{res}[\bar{P}_j(\zeta)]\rightarrow \sum_q\bar{res}[\bar{P}_j
\bar{\partial}^q{(\zeta-z)^q \over q!}]
$$
{\tenbf Proof.} From (5.6) one can recognize that if right
derivations act on the coefficients $a_i(\zeta)$ in $Y$ then
right residues becomes zero such as under the definition
there are no right integral symbols in
$\bar{Y}$ and $\bar{L}(\partial+{1\over k}\bar{\partial}-\phi'_1)^{-1}$.
Therefore, functions $a_i(\zeta)$ do not take part
in the ordering in accordance with the procedure
represented in the section 3. Q.E.D.

Now we are capable to complete computation of the
quantum corrections:
$$
\eqalign{
&:\bar{res}
{(\bar{Y}_z\bar{\partial}^{-1}_z\bar{L}^\phi_z)}_{+}F_s(z):=\cr
&=:(\hat{Y}\hat{L}^{\phi})_{+}F_s):-:{1\over k}\sum_{jql}\bar{res}[C^j_l
C^{j-l}_q\bar{P}_j(z)\bar{\partial}^{l+q+1}
{(-1)^l \over l+q+1}]
F_s^{(j-l-q)}:\cr
&=:(\hat{Y}\hat{L}^{\phi})_{+}F_s):
-:\sum_j\bar{res}[{\bar{P}_j \bar{\partial}\over k}F_s^{(j)}]: \quad.
\cr}$$
Taking into account eq.(5.9) one can see that the
following equality is obeyed
$$
-:\bar{res}[{(\bar{Y}\bar{\partial}^{-1}\bar{L})}_{+} F]:=
-:\bar{res}[\bar{P}_{-1})]F_s(z): \quad .
$$
These together with (5.7) prove (5.4) i.e.
$$
\delta_HF_s=-:\bar{res}
[{(\bar{Y}\bar{\partial}^{-1}\bar{L})}_{+} F_s]:=-:(\hat{Y}\hat{L})_{+}
F_s: \quad Q.E.D.  \quad.
$$
\vglue 12pt
\line{\tenbf 5.CONCLUSION. \hfil}
\vglue 5pt
Thus we have used right derivation technik to prove that
the vertex operators $F_s$ $(s=1 ,..., n)$ (3.6) are transformed
under the $W$ algebra action as
$$
\delta_HF_s=-:[{(\hat{Y}\hat{L})}_{+} F_s]:   \quad .
$$
Another rigorous result is that these operators must to
satisfy the "null-vector equation":
$$
:[\hat{L}F_s]:=0 \quad .
$$
We have seen above that provided choose a
suitable deformation of the (pseudo)\-differen\-tial
operators
these formulae and quantum Miura
transformation (3.16)
for W-algebras have precisely the same form as
classical ones. It is worth noting that
our deformation is independent of the n.

It will be very important to develop these results
for the $W$-algebras associated to $SL(n)$ group.
Another vital problem is the finding of the quantum
analogue of the Gel'fand-Dikii bracket in terms
of pseudodifferential operators. Unfortunately,
direct application of isomonodromic deformation
method proposed in [23] gives unproper result
becouse it is necessary to have into account
some subtle problems with normal ordering [15].
\vglue 12pt
\line{\tenbf ACKNOWLEDGMENTS \hfil}
\vglue 5pt
It is a pleasure to thank S.Lukyanov for suggesting
this problem to me
and enlightening discussions. I also thank
A.Belavin, B.L.Feigin, M.Lashkevich, S.Par\-cho\-men\-ko
and H.Thoomasyan for stimulating discussions.

\vglue 12pt
\line{\tenbf References \hfil}
\vglue 5pt

\medskip
\ninerm
\baselineskip=11pt
\frenchspacing

\item{1} A.B.Zamolodchikov, {\nineit Theor. Math. phys.} {\ninebf 65},
         1205 (1986).
\item{2} A.B.Zamolodchikov and V.A.Fateev, {\nineit Nucl.Phys.} {\ninebf B280
        [FS18]}, 644 (1987).
\item{3} M.Gel'fand and L.A.Dikii, preprint {\nineit IPM AN SSSR}
        {\ninebf 136}, (1978), Moscow.
\item{4} A.A.Belavin,A.M.Polyakov,A.B.Zamolodchikov, {\nineit Nucl.Phys.}
{\ninebf B241}, 333 (1984).
\item{5} S.Lukyanov, {\nineit Funct.Anal.Appl.} {\ninebf 12}, 466 (1988).
\item{6} V.A.Fateev,S.L.Lukyanov, {\nineit Intern.J.Mod.Phys.}
         {\ninebf A7}, 853 (1992);
\item{ } {\ninebf A7}, 1325 (1992).
\item{7} V.A.Fateev,S.L.Lukyanov, {\nineit Intern.J.Mod.Phys.}
       {\ninebf A3}, 507 (1988).
\item{9} A.Bilal and J.-L.Gervais, {\nineit Phys.Lett.}
         {\ninebf B206}, 412 (1988).
\item{10} O.Babelon, {\nineit Phys.Lett.} {\ninebf B215}, 523 (1988).
\item{11} M.Bershadsky, H.Ooguri, {\nineit Commun. Math.Phys.}
        {\ninebf 126}, 49 (1989).
\item{12} B.Feigin,E.Frenkel, {\nineit Phys.Lett.} {\ninebf B240}, 75 (1990).
\item{13} B.Feigin,E.Frenkel,Preprint {{\nineit MSRI} {\ninebf 04029-91}
\item{14} A.A.Belavin {\nineit Adv.Stud.Pure Math.} {\ninebf 19}, 117 (1989).
\item{15} F.Bais,P.Bowknegt,K.Schoutens,M.Surridge, {\nineit Nucl.Phys.}
 {\ninebf B304}, 348 (1988).
\item{16} V.G.Drinfel'd,V.V.Sokolov, {\nineit "Mod.probl.in math."}
        Moscow VINITI, 81 (1984).
\item{17} P.Matheu, {\nineit Phys.Lett.} {\ninebf B208}, 412 (1988).
\item{18} P.Di Francesco, C.Itzykson and J.-B.Zuber,
{\nineit Sacley preprint } {\ninebf SPHT/90-149}
\item{19} V.Yu.Ovsienko, B.A.Khesin, {\nineit Funct.Anal.Appl.}
         {\ninebf 24}, 33 (1988).
\item{20} B.A.Khesin, {\nineit Comm.Math.Phys.}, {\ninebf 145}, 357 (1992).
\item{21} M.Adler, {\nineit Inv.Math.} {\ninebf 50}, 219 (1979).
\item{22} J.Felder, {\nineit Nucl.Phys.} {\ninebf B317}, 215 (1989).
\item{23} Ya.P.Pugay, {\nineit Phys.Lett.} {\ninebf B279}, 34 (1992).
\item{24} B.Khesin, I.Zakharevich, Preprint {{\nineit IHES} (1992).
\item{25} B.Enriquez, Preprint (1992).

\vfil\supereject
\bye